\documentclass[runningheads]{svmult}

\usepackage{makeidx}   
\usepackage{graphicx}  
\usepackage{subeqnar}  
\usepackage{multicol}  
\usepackage{physprbb}  
\makeindex             

\usepackage{amsmath}   

\newcommand{\be}{\begin{equation}}
\newcommand{\ee}{\end{equation}}
\newcommand{\bea}{\begin{eqnarray}}
\newcommand{\eea}{\end{eqnarray}}

\begin{document}



\title*{Quenching Effects in the Hadron Spectrum}
\toctitle{Quenching Effects in Lattice Hadron Physics}
%
%
\titlerunning{Unquenching Effects in Lattice Hadron Physics}
%
\author{C. Allton}
\authorrunning{C. Allton}
%
%
\institute{
Department of Physics,
University of Wales Swansea,
Swansea SA2 8PP,
U.K.}

\maketitle              

\begin{abstract}
Lattice QCD has generated a wealth of data in hadronic physics over the
last two decades. Until relatively recently, most of this information
has been within the \lq\lq quenched approximation'' where virtual
quark--anti-quark pairs are neglected.
This review presents a descriptive discussion of the effects
of removing this approximation in the calculation of hadronic masses.
\end{abstract}



\section{The Quenched Approximation}
\label{sec:intro}

In a quantum field theory involving gauge and fermion degrees of freedom,
such as QCD, we have the following path integral formalism for
the expectation value of a quantity $\Omega$:
\be
\langle \Omega \rangle = \frac{1}{Z} \int {\cal D} \psi {\cal D} \bar{\psi} {\cal D} A \; \Omega (\psi ,\bar{\psi} ,A) \; \E^{-S_E(\psi ,\bar{\psi} ,A)}
\ee
where $Z$ is the usual path integral and the Euclidean action $S_E$ is
defined in terms of the usual field strength tensor $F_{\mu\nu}$,
\be
S_E = \int d^4x \left\{ \bar{\psi}(x)(D \!\!\!\! / +m) \psi(x)
+ \frac{1}{4} F_{\mu \nu} F_{\mu \nu} \right\}.
\label{eq:contaction}
\ee
The gauge degrees of freedom, $A$, are bosonic, but the fermionic degrees
of freedom, $\psi$ are fermionic and hence anti-commute. These are difficult
to deal with in a computer simulation, but fortunately, since they occur as
they can be integrated $\psi$ analytically resulting in the usual determinant
factor:
\be
\langle \Omega \rangle = \frac{1}{Z} \int {\cal D} A \;
\Omega (\psi ,\bar{\psi} ,A) det(D \!\!\!\! / + m) \; \E^{-S_E(\psi ,\bar{\psi} ,A)}.
\ee

Simulations of this quantum field theory are performed on a space--time lattice
by simply replacing all the continuous derivatives and integrals with
finite differences and sums over gauge configurations. Hence, we have
on the lattice:
\be
\langle \Omega \rangle = \frac{1}{Z} \sum_{\{U\}} \Omega(\psi ,\bar{\psi} ,A)\;
\E^{-S_g} \,det(\Delta \!\!\!\! / + m)
\label{eq:lat}
\ee
The (naive) lattice Euclidean action is
\be
S = S_F + S_g =
\sum_x \left\{ \bar{\psi} (\Delta \!\!\!\! / + m) \psi(x) \right\} - \frac{1}{g_0^2} \sum_p Tr(U_p+U_p^\dagger)
\label{eq:lataction}
\ee
where $a$ is the lattice spacing,
the link variable $U_\mu(x)$ now carries the gauge degrees of freedom,
and $U_p$ is the trace of the product of link variables around a plaquette,
\be
  U_p = U_\mu (x) U_\nu (x + \hat{\mu} a) U^\dagger_\mu (x + \hat{\nu} a) U^\dagger_\nu (x).
\ee
This formalism maintains gauge invariance even on a lattice \cite{wilson}.

Variations of the naive lattice action can be made to improve its 
convergence to the continuum action in two areas:
\begin{itemize}
\item the naive action suffers from fermion doubling -- each lattice quark
flavour corresponds to $2^d$ continuum flavours, where $d$ is the space--time
dimension;
\item lattice actions in general suffer from discretisation errors which
enter when the continuum derivatives in (\ref{eq:contaction}) is replaced
by the finite difference in (\ref{eq:lataction}).
\end{itemize}
Two methods are generally used to overcome the first difficulty -- the
Wilson/clover family of actions, and the staggered action.
Both of these actions can be tweaked so that their lattice systematic
error (the second difficulty above) are reduced, and then they are termed
``improved''.

Simulations using the lattice formalism can be performed by replacing
the naive sum in (\ref{eq:lat}) with a Monte Carlo estimate. This
introduces a statistical error ${\cal O}(1/\sqrt{N_{cfg}})$ in the
estimate of $\langle \Omega \rangle$ where $N_{cfg}$ is the number of
configurations in the Monte Carlo sum.

The lattice prescription of formulating a Quantum Field Theory has \'a
priori no model assumptions --
it is derived exactly from the full continuum formalism with no
approximations.
However the parameter values in real
computer simulations of lattice QCD are far from their
experimental values.  This is due to limitations in current computer
power! Table \ref{tab:approx} lists the
values of the parameters in typical lattice simulations along with
their experimental values.
Thus typical lattice simulations must inevitably rely on some
extrapolation of lattice data.
Note that the bare lattice gauge coupling, $g_0$, is not listed
in table \ref{tab:approx}. This because the information about $g_0$
is contained within the $a$ value, through dimensional transmutation.
Our usual intuition about high momentum transfers (short--distance
physics) corresponding to the weakly coupled regime (small values
of $g_0$) in asymptotically free theories such as QCD, is directly
applicable to lattice simulations. So we have $g_0\rightarrow 0$
as $a \rightarrow 0$.


\begin{table}
\caption{Typical parameter values in current lattice simulations}
\begin{center}
\renewcommand{\arraystretch}{1.4}
\setlength\tabcolsep{5pt}
\begin{tabular}{lll}
\hline\noalign{\smallskip}
parameter & typical value & experimental value \\
\noalign{\smallskip}
\hline
\noalign{\smallskip}
$m_q$     & 
{\raisebox{-.5ex}{\rlap{$\,\,\sim$}} \raisebox{.7ex}{$>$}}
$m_s/2 \approx 50\,MeV$ & $m_u, m_d \approx 5\, MeV$ \\
$a$       & $0.05 - 0.20\,fm$ & 0 \\
$L$       & $2-4 \,fm$ & $\infty$ \\
$N_{cfg}$ & ${\cal O}(100)$ & $\infty$ \\
$N_f$     & $0, 2$ or $2+1$ & \lq\lq 2+1'' \\
\hline
\end{tabular}
\end{center}
\label{tab:approx}
\end{table}

\begin{figure}[b]
\vspace*{10mm}
\begin{center}
\includegraphics[width=.8\textwidth]{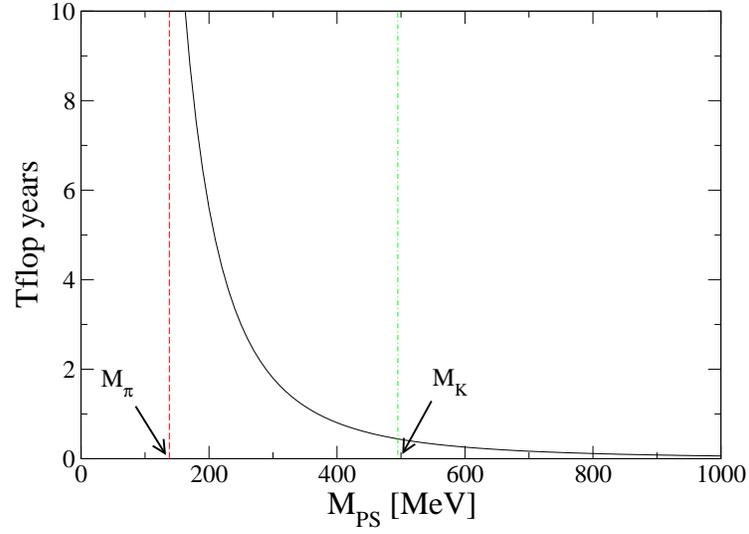}
\end{center}
\caption[]{
The computer time in Teraflop--years required for a full lattice
QCD simulation as a function of pseudoscalar meson mass using
the formula for clover actions in \cite{hartmut}. 
We have assumed
(i) a lattice spacing of $a=0.1\,fm$;
(ii) a lattice volume of $(3\,fm)^4$; and
(iii) that 200 configurations in the ensemble sum in (\ref{eq:lat})
are required.
The physical points corresponding to the $\pi-$ and K-mesons are shown
by vertical lines.
}
\label{fig:cost}
\end{figure}

Equations (\ref{eq:lat} \& \ref{eq:lataction}) correctly define the full
continuum theory in the limit as the lattice spacing $a\rightarrow 0$.
However, it is
extremely expensive to simulate with (\ref{eq:lat} \& \ref{eq:lataction}).
Figure \ref{fig:cost} shows the estimated cost of lattice calculations
as a function of quark mass using the formula in \cite{hartmut} for the
``clover'' action.
(Actually, the horizontal axis of this plot is $M_{PS}$, but, from the PCAC
relation, we have $M_{PS} \propto \sqrt{m_q}$.)
In Fig. \ref{fig:cost} we have assumed: (i) a lattice spacing of $a=0.1\,fm$;
(ii) a lattice volume of $L^4 = (3\,fm)^4$; and (iii) that there are
$N_{cfg}=200$ configurations in the ensemble sum in (\ref{eq:lat}).
(These are very conservative assumptions!)
As can be seen, for even modest values of $M_{PS} \approx \frac{1}{2}M_K
\sim 250\,MeV$, full simulations require Tera-scale computing.
\footnote{Recent advances in lattice actions, e.g. using an improved
staggered action, have meant that CPU
requirements are not quite so pessimistic \cite{breakthrough}.}


For this reason, \cite{quenched} introduced the \lq\lq quenched''
approximation where the true QCD vacuum is replaced with one
with no quarks present (i.e. $N_f=0$ in Table \ref{tab:approx}).
Specifically the quenched approximation is defined as follows:
\begin{itemize}
\item $det(\Delta \!\!\!\! / + m)$ is replaced by unity, thereby
removing the quark--anti-quark loops from the vacuum configurations;
\item the coupling $\beta=6/g_0^2$ is shifted to try to counteract
(as much as is possible) the removal of these $q-\bar{q}$ pairs. Typically
this shift is of the following order $\beta^{Q} \approx \beta^{full} + 0.6$
where $\beta^{Q,full}$ refer to the coupling in the quenched and full
theories.
\end{itemize}

\begin{figure}[b]
\vspace*{10mm}
\begin{center}
\includegraphics[width=.8\textwidth]{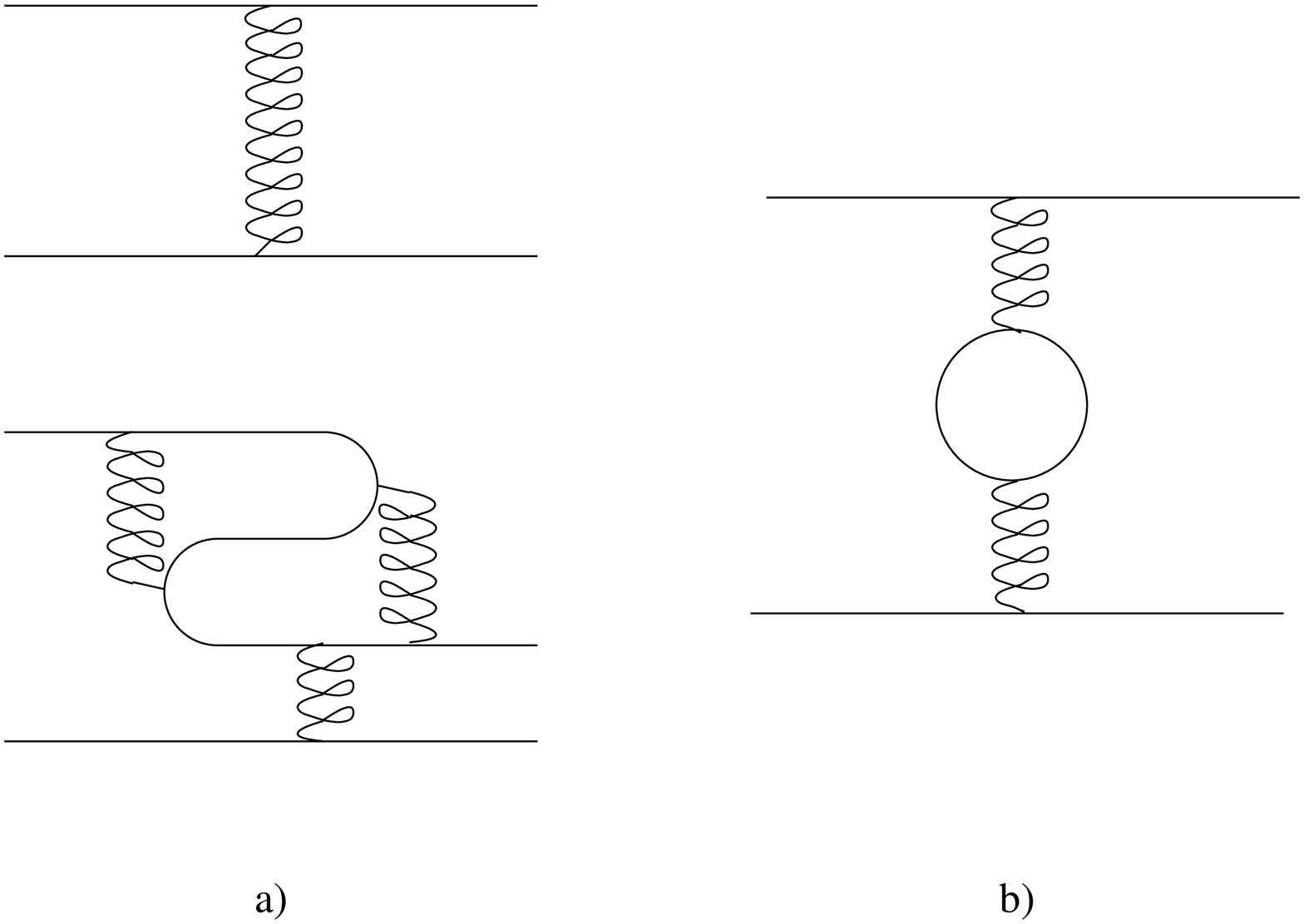}
\end{center}
\caption[]{
Diagrams which are present in (a) both quenched QCD and full QCD, and
(b) present only in full QCD.
The full lines are quarks and the spirals are gluons.
}
\label{fig:feynman}
\end{figure}

In this way the quenched approximation can be viewed as an
effective field theory, i.e. it contains a subset of all 
the interactions, and the couplings of the quenched theory have
to be tuned to take care of these missing diagrams.
Figure \ref{fig:feynman} shows the diagrams which are present in both the
quenched and full theory, and those which are present only in the full theory.


Quenched simulations are several orders of magnitude faster than
full (unquenched) simulations, and full simulations have only been
performed in earnest in the last 5 years or so.
Typical statistical and systematic errors of state--of--the--art
full simulations are of the same order now as quenched simulations' errors
were a decade ago. Particle physics is primarily concerned with the
comparison of theory with experiment, and when theoretical calculations
have inherent errors, it is crucial to understand and quantify their scale.

The main aim of this chapter is to determine the systematic effect introduced
in the hadron spectrum by the quenched approximation.
We will find that uncovering quenching effects is more difficult than
one would first imagine for two reasons:
\begin{itemize}
\item the quenched approximation proves to be surprisingly successful
for many hadronic quantities, i.e. it  reproduces much of the hadron
spectrum at the 5--10\% level. Assuming that QCD is the theory of the
strong interaction implies that removing this approximation makes
a relatively small effect!
\item current full simulations have statistical errors of a few percent
(since they are highly cpu-intensive) and so discerning the
quenching effects with this relatively noisy data can prove difficult;
\end{itemize}

While the quenched approximation proves to be unintuitively accurate
for many hadronic quantities, there are some quantities where it
either fails drastically, or has pathologies when the valence quark masses
in the hadrons becomes vanishing. Examples of this include:
\begin{itemize}
\item the $\eta$ and $\eta'$ mesons are degenerate in the quenched theory,
whereas they {\em not} degenerate in full QCD.
This is because the quenched theory excludes diagrams involving disconnected
$q-\bar{q}$ loops. (See \cite{cmi_eta} for a description of lattice
simulations of $\eta$ and $\eta'$.)
\item the chiral limit of quenched QCD suffers from ``chiral logs''
$\sim \log(m_q/\Lambda_\chi)$ where $m_q$ is the quark mass and $\Lambda_\chi$
is a mass parameter proportional to pion decay constant.
These logarithms enter the chiral perturbation theory expansion of
various hadronic masses in the quenched theory, spoiling their chiral
limit \cite{qqcd_chiral_expansion}
\item the hyperfine mass splittings in heavy--mesons in the quenched theory
is wrong by up to 10\% or more, with the sign of the discrepancy depending
on the states considered. This systematic error is greatly reduced when
the full theory is considered.
\end{itemize}
We will discuss some of these issues in later sections.

The next section briefly reviews the best current results obtained from
the quenched approximation. It outlines the accuracy of this approximation
for the hadronic spectrum. Section \ref{sec:full} presents
recent results from full (i.e. unquenched) simulations and
we attempt to uncover estimates of quenching effects in Sect. \ref{sec:diff}.



\section{Results from the Quenched Approximation}
\label{sec:quenched}

While there have been many papers published using lattice simulations
in the quenched approximation\footnote{ A search on the SPIRES
database for \lq\lq quenched'' returns more than 500 papers, and this
does not include papers which use quenched simulations but where the
authors have not included this word in the paper's title!}
we will concentrate on the work of the CP-PACS collaboration who have
produced one of the most accurate quenched study of the light hadron
spectrum in
\cite{cppacs}\footnote{Another paper from the CP-PACS Collaboration studies
even larger lattice of up to $64^3 \times 112$ but the lattice action
employed in this work is the pure Wilson action which has ${\cal O}(a)$
errors \cite{cppacsq}.}.
Their calculations used an improved clover action \cite{clover}
simulated at volumes of around $2.5^3$ fm$^3$ with
several quark masses and lattice spacings.
They are thus able to perform continuum ($a\rightarrow 0$)
and chiral ($m_q \rightarrow m_{u,d}$) extrapolations (see Table
\ref{tab:approx}).


\begin{figure}[b]
\begin{center}
\includegraphics[width=.8\textwidth]{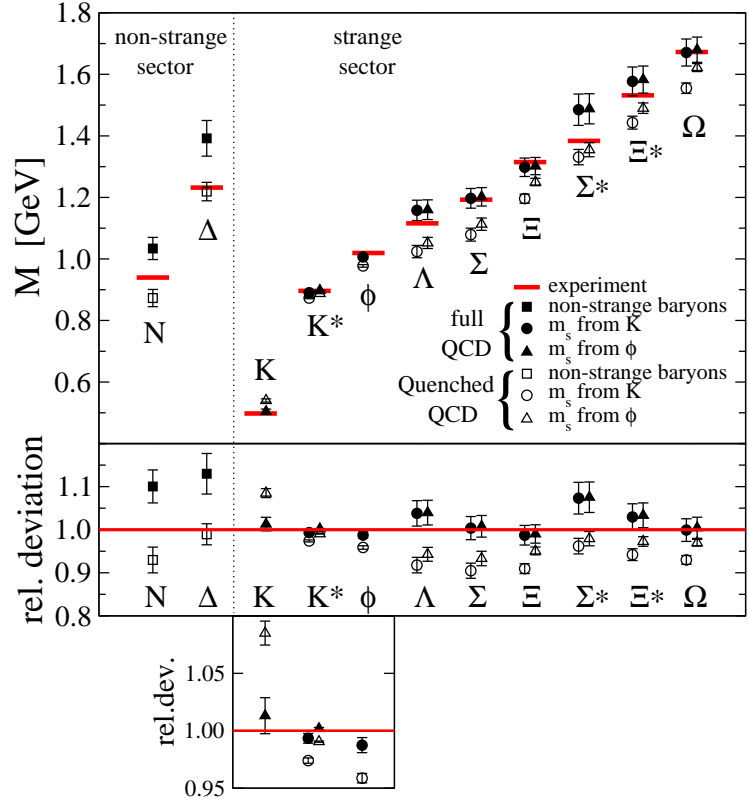}
\end{center}
\caption[]{
The light hadron spectrum from CP-PACS \cite{cppacs},
tables XV and XII.
Both the quenched and full QCD results are shown, together with
the experimental value.
The two hadrons on the left do not contain strange quarks,
whereas the other hadrons do.
The lattice spacing was set from the $\rho-$mass.
Two methods were used to set the strange quark mass:
from (i) the $K$ and (ii) the $\phi-$mass.
The results of both definitions are shown.
In the middle and lower plots, the relative deviation
($= M_{lattice}/M_{expt}$) is shown.
The lower plot is a close-up of the middle plot showing
the relative deviation for the strange mesons.
}
\label{fig:cppacs}
\end{figure}


An impressive summary of their quenched spectrum for the light hadrons
is shown in Fig. \ref{fig:cppacs} as the open symbols.  Their lattice
data are shown {\em after} the appropriate continuum and chiral
extrapolations and is taken from table XV of \cite{cppacs}.  As can be
seen from the middle panel of Fig. \ref{fig:cppacs}, discrepancies
between the quenched lattice results and the experimental values are
${\cal O}(5-10\%)$.  It is important to note that this relatively
small difference is only discernible due to the tiny errors in the
lattice data of ${\cal O}(1-2\%)$.  Quenched calculations of an
earlier generation \cite{weingarten}, with correspondingly larger
errors, were not able to uncover deviations from experimental values.

Figure \ref{fig:cppacs} contains two sets of lattice data points:
those obtained from the $K$ and $\phi$ inputs. These refer to the
hadron whose mass was used to set the strange quark mass in the
lattice calculation. (The $\rho$ mass was used to define the lattice
spacing, $a$.) The fact that there are differences between
these two sets of data is itself an indication of the failure of the
quenched approximation, i.e. an exact calculation's result wouldn't
depend on how the scale was set.

The CP-PACS collaboration also find the mass splittings, such as the
hyperfine splitting in the meson sector and the splittings in the
decuplet (baryon) sector are smaller than experimental values.

A further indication of the failure of the quenched approximation is
in the determination of the strange quark mass $m_s$. As mentioned
above, this quantity can again be calculated using either the $K-$ or
$\phi-$meson as input, but the deviation between the two results is at
the $3-4\sigma$ level.

Moving to the heavy--hadron mass spectrum, a recent publication, using
an improved staggered action, studied the splittings in the heavy--meson sector
\cite{breakthrough}.  The left--hand plot in Fig. \ref{fig:heavysplit}
(taken from \cite{breakthrough}) shows the quenched predictions of various
splittings from \cite{breakthrough}. This clearly shows a discrepancy
between the quenched results and experimental values.
As we will see in Sect. \ref{sec:fullheavy}, this discrepancy disappears
when we remove the quenched approximation.



\section{Results from Full (unquenched) Simulations}
\label{sec:full}


This section will give a flavour of current full QCD lattice simulations
by concentrating on the CP-PACS \cite{cppacs} and UKQCD \cite{ukqcd}
collaborations' results for the light hadron spectrum, and the
work of \cite{breakthrough} for the heavy--meson spectrum.
Both the CP-PACS and UKQCD collaborations used two flavours of
improved clover fermions whereas the collaboration in \cite{breakthrough}
used an improved staggered action (which has a cpu advantage over the
Wilson action, see Sect. \ref{sec:intro} and \cite{panel}).

Table \ref{tab:full} displays the parameters used in these
collaborations' work.
Note that we have differentiated the {\em sea} and {\em valence}
quark masses in this table (c.f. table \ref{tab:approx}).
The sea quarks are those which always appears in quark loops and are
{\em not} connected to the source/sink operators (e.g. the quark
loop in Fig. \ref{fig:feynman}), and the valence quarks are those
which enter the source/sink interpolating operators.

As can be seen from table \ref{tab:full}, the CP-PACS collaboration
have performed QCD simulations at parameter values closer to the
experimental values and has larger statistics than the UKQCD
collaboration (see also Table \ref{tab:approx}).
However the UKQCD collaboration chose a subset of
its parameter values so that the lattice spacing remained fixed as the
sea quark mass, $m_q^{sea}$, varied. This meant that ${\cal O}(a)$ effects
could more readily be disentangled from dynamical quark effects.
Furthermore, the UKQCD lattice action has the technical advantage that
it has no ${\cal O}(a)$ lattice systematic errors.

In the simulations of \cite{breakthrough} extremely light quarks were
able to be studied due to the use of the improved staggered action.
(This seems to have become the action of choice for most dynamical
simulations.)  Also \cite{breakthrough} simulates with the more
physical value of 2+1 quark flavours (see table \ref{tab:approx}).


\begin{table}
\caption{Indicative parameter values used in full QCD simulations
by the CP-PACS \cite{cppacs} and UKQCD Collaborations \cite{ukqcd}
in their study of light hadrons, and in the study of the heavy--meson spectrum
in \cite{breakthrough}.
}
\begin{center}
\renewcommand{\arraystretch}{1.4}
\setlength\tabcolsep{5pt}
\begin{tabular}{llll}
\hline\noalign{\smallskip}
parameter & CP-PACS \cite{cppacs} & UKQCD \cite{ukqcd}
          & Davies et al. \cite{breakthrough}\\
\noalign{\smallskip}
\hline
\noalign{\smallskip}
$m_q^{sea}$ & $0.5 m_s - 1.8 m_s$             
                                   & $0.6 m_s - 2.0 m_s$             
                                                     & $0.17 m_s - 0.5 m_s$\\ 
            & i.e. ${\cal O}(50-180)$ MeV    
                                   & i.e. ${\cal O}(60-200)$ MeV   
                                                     & i.e. ${\cal O}(17 - 50)$ MeV\\ 
$m_q^{val}$ & $0.25 m_s - 2.1 m_s$             
                                   & $0.6 m_s - 2.4 m_s$             
                                                     & $0.12 m_s - m_s$\\ 
            & i.e. ${\cal O}(25-210)$ MeV   
                                   & i.e. ${\cal O}(60-240)$ MeV   
                                                     & i.e. ${\cal O}(12 - 100)$ MeV \\ 
$a$       & $0.09 - 0.25\,$fm & $\sim 0.11\,$fm & $0.09$ fm \&  $0.12$ fm\\
$L$       & $\sim 2.5\,fm$    & $\sim 1.7 \,fm$ & $\sim 2.5$ fm \\
$N_{cfg}$ & ${\cal O}(1000)$  & ${\cal O}(200)$ &  \\
$N_f$     & $2$               & $2$             & $2+1$ \\
\hline
\end{tabular}
\end{center}
\label{tab:full}
\end{table}


Rather than give the full details of the results from these
collaboration's work, a summary is presented in the following.  In the
next section we attempt to understand the discrepancies between this
section's full QCD results and quenched simulations from
Sect. \ref{sec:quenched}



\subsection{Meson spectrum}

\begin{figure}[b]
\begin{center}
\includegraphics[width=.8\textwidth]{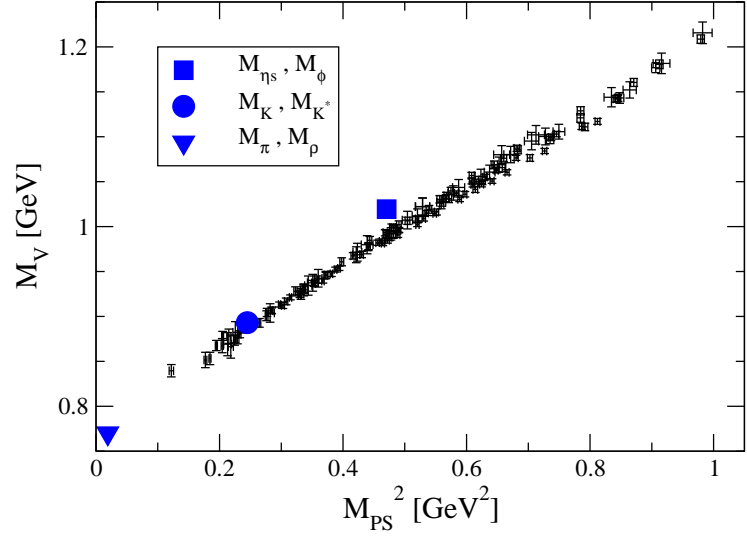}
\end{center}
\caption[]{
The light meson masses from CP-PACS \cite{cppacs} (see also \cite{wes}).
In this figure, we have set the lattice spacing from
the method described in \cite{leonardo}.
Also shown are the experimental points.
}
\label{fig:cppacs_meson}
\end{figure}

In Fig. \ref{fig:cppacs_meson} we plot the vector and pseudoscalar
meson mass taken from \cite{cppacs} together with the experimental
points.  In this figure, the lattice spacing, $a$, was determined from
the $K$ and $K^\ast$ meson masses using the method described in
\cite{leonardo}. The huge number of data points corresponds to the
fact that there are 16 different ($\beta$,$m_q^{sea}$) combinations in
\cite{cppacs} and that there are 9 different valence quark
combinations for each of these ($\beta$,$m_q^{sea}$) ensembles.  Also
plotted are the experimental data points corresponding to the
$(\pi,\rho)$, $(K,K^\ast)$, and $(\eta_s,\phi).$\footnote{ Note that
$\eta_s$ is not a physical particle, since there is no pure
$s-\bar{s}$ pseudoscalar meson due to mixing with the $u,d$ quarks.
The $\eta_s$ mass shown here is defined as $M_{\eta_ s}^2 = 2 M_K^2 -
M_\pi^2$ which follows from the PCAC relationship $M_{PS}^2 \propto m_q$.}

One of the main points to be taken from this plot is that
the systematics involved in lattice simulations are clearly under control.
Variations amongst this data in Fig. \ref{fig:cppacs_meson} is 
\;\raisebox{-.5ex}{\rlap{$\sim$}} \raisebox{.5ex}{$\!\!<$}\;1\%,
while the lattice spacing and sea quark mass vary by around a factor of three:
$a \sim 0.09 - 0.25\,$fm, and $m_q^{sea} \sim 0.5 - 1.8 m_s$.
A close analysis of the data has been used to extrapolate away these residual
systematic effects in $m_q^{sea}$ and $a$ \cite{cppacs} (see also \cite{wes}).

Figure \ref{fig:cppacs} also shows the hadronic spectrum, including the
three mesons, $K, K^\ast$ and $\phi$ taken from \cite{cppacs}.
These have been obtained by chirally extrapolating data analogous to
that in Fig. \ref{fig:cppacs_meson} to the physical quark masses.
As can be seen (particularly in the
lower panel of Fig. \ref{fig:cppacs}) the full QCD simulated results are
in very good agreement with experiment.

As mentioned above, the $K,K^\ast$ mass is used to set the lattice
spacing, $a$, in Fig. \ref{fig:cppacs_meson} \cite{leonardo}.  This
means that the lattice data and the experimental $K,K^\ast$ point agree by
construction. 
However, the {\em slope} of the lattice data is a real lattice prediction.
In the next subsection we study this gradient.


\subsection{$J-$parameter}

In this section we analyse the gradient $dM_V/dM_{PS}^2$ of the lattice data.
The dimensionless quantity used to study this is defined \cite{cmi}
\be
\label{eq:J}
J = M_V \frac{dM_V}{dM_{PS}^2} \bigg|_{K,K^\ast}.
\ee
Note that we define the {\em experimental} value of the
$J-$parameter by approximating the derivative in (\ref{eq:J})
by a finite difference:
\be
\label{eq:J_discrete}
J^{discrete} = M_{K^\ast} \frac{M_{K^\ast}-M_\rho}{M_K^2-M_\pi^2}.
\ee
Therefore the lattice estimate of $J$ is obtained by taking the
derivative in (\ref{eq:J}) {\em w.r.t.} variations in
$M_{PS}^2(m_q^{sea},m_q^{val})$ at {\em fixed} $m_q^{sea}$
but varying {\em valence} quark mass, $m_q^{val}$
(i.e. the experimental/physical sea quark mass is clearly fixed!).

\begin{figure}[b]
\begin{center}
\includegraphics[width=.8\textwidth]{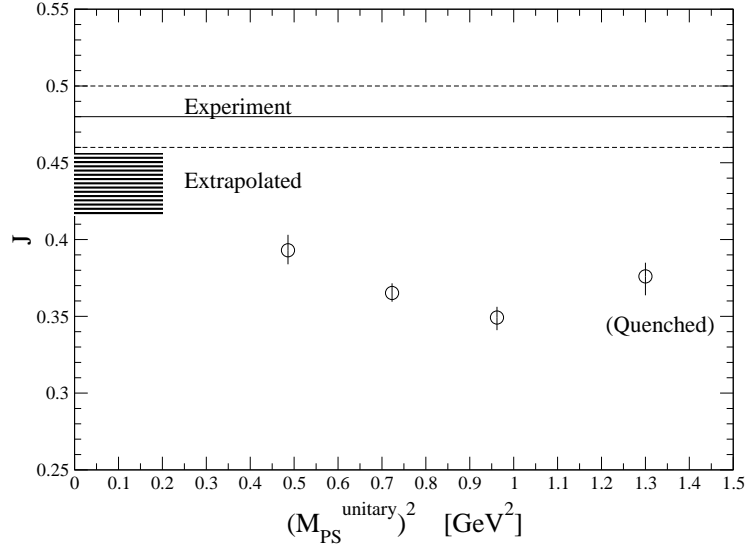}
\end{center}
\caption[]{
$J$ versus $(M_{PS}^{unitary})^2$ using the approaches described in the text
from \cite{ukqcd}.
$M_{PS}^{unitary}$ is the pseudoscalar meson mass comprising of degenerate
valence quarks which are themselves degenerate with the sea quarks.
(Note, from PCAC, $(M_{PS}^{unitary})^2 \propto m_q^{sea}$.)
The quenched data points have been plotted at
$(M_{PS}^{unitary})^2 = 1.3\,GeV^2$ for convenience.
The banded region at the left of the graph is the result
of the extrapolation $m_q^{sea} \rightarrow 0$ for the full QCD data.
The experimental value of $J = 0.48(2)$ is also shown.
}
\label{fig:ukqcd_j}
\end{figure}

The $J-$parameter has been studied in both \cite{cppacs} and \cite{ukqcd},
but we concentrate here on the analysis in \cite{ukqcd}.
Figure \ref{fig:ukqcd_j} plots the $J-$parameter from \cite{ukqcd}.
The $J-$parameter is calculated at each of three $m_q^{sea}$ values
separately.
These $J(m_q^{sea})$ values are then extrapolated in $m_q^{sea} \propto
M_{PS}^2$ to the physical point $m_q^{sea} \approx 0$.
This extrapolated $J$ value is shown as a banded region in
Fig. \ref{fig:ukqcd_j}.


As can be seen from Fig. \ref{fig:ukqcd_j} the individual $J(m_q^{sea})$ values
are significantly smaller than the experimental value. However there is a
clear trend in $m_q^{sea}$ which tends towards the experimental point.



\subsection{Baryons}
\label{sec:baryons}

\begin{figure}[b]
\begin{center}
\includegraphics[width=.8\textwidth]{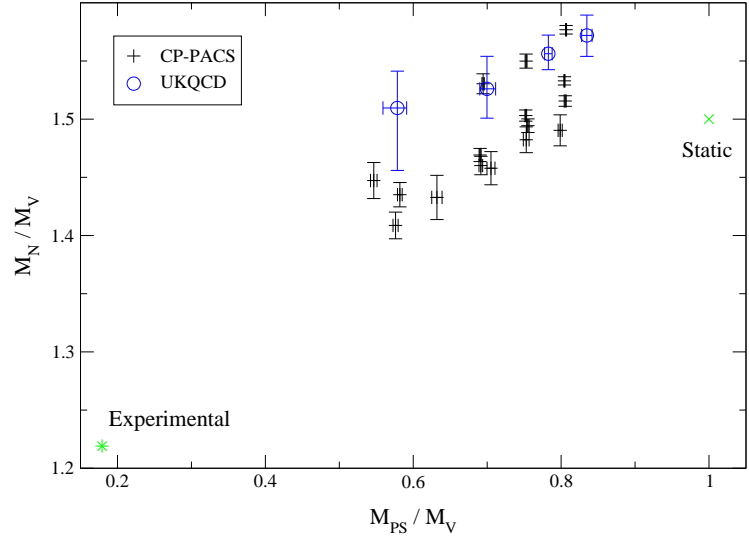}
\end{center}
\caption[]{ The \lq\lq Edinburgh plot'' for selected full QCD data
from \cite{cppacs,ukqcd}.  The lattice data points shown are the {\em
unitary} points (i.e. $m_q^{valence} \equiv m_q^{sea}$).  The
experimental point is shown, along with the static limit ($m_q
\rightarrow \infty$).  }
\label{fig:edplot}
\end{figure}

The vector and pseudoscalar light--meson sector at various quark masses
is defined by the plot in Fig. \ref{fig:cppacs_meson}.
Traditionally, the corresponding plot containing information
about the light--baryon sector (specifically the nucleon) is the
``Edinburgh'' plot where the nucleon mass is plotted against the 
pion mass (with both masses normalised by the vector meson mass).
Figure \ref{fig:edplot} shows this plot for the
the CP-PACS \cite{cppacs} and UKQCD collaborations
for their {\em unitary} data, i.e. where $m_q^{valence} \equiv m_q^{sea}$.
As can be seen, there is a relatively large spread in the data,
but there is a tendency  for the data to approach the experimental
point as the quark masses decrease.

After this chiral extrapolation is performed, the CP-PACS collaboration
\cite{cppacs} obtained the baryonic spectrum seen in Fig. \ref{fig:cppacs}.
This is an impressive array of data spanning octet and decuplet sectors.
As can be seen Fig. \ref{fig:cppacs} the nucleon and $\Delta$ differ from
experiment by around 10\%, whereas the $\Xi, \Xi^\ast$
(with quark content $lss$)
and particularly the $\Omega$ (with quark content $sss$) are in perfect
agreement with experiment.
This implies that lattice simulations become more accurate
as the strange quark content increases \cite{cppacs}.

Since lattice simulations normally have valence quarks which span the
mass of the strange (see table \ref{tab:full}), the spectrum calculation
of baryons containing purely strange valence quarks requires no valence quark
chiral extrapolation.
However, the level of chiral extrapolation required obviously becomes
more and more significant as the light quark (i.e. $u$ and $d$) content
of the baryon increases.
This suggests that chiral extrapolation procedures need
more careful consideration in order to resolve the discrepancy above
(see \cite{adelaide_chiral}).
Note that the authors of \cite{cppacs} themselves argue that this discrepancy
could be due to finite volume effects which impinge upon baryons composed
of light quarks more than those composed of strange quarks.
This could presumably also be a factor, particularly since finite volume
effects are likely to {\em increase} the mass (which is in the direction
of the observed discrepancy in Fig. \ref{fig:cppacs}) and be most relevant
for the lightest baryons.


\subsection{Heavy--Meson Mass Splittings}
\label{sec:fullheavy}

There has been a recent study of the heavy--meson spectrum in
\cite{breakthrough} which uses 2+1 flavours of quarks, i.e. 2 light degenerate
flavours which play the role of the $u$ and $d$ quarks, and one heavier
(but still dynamical) quark which plays the role of the $s$ quark.
This is obviously closer to the real world than the simulations of
the CP-PACS and UKQCD collaborations (see tables \ref{tab:approx} \&
\ref{tab:full}).

\begin{figure}[b]
\begin{center}
\includegraphics[width=.8\textwidth]{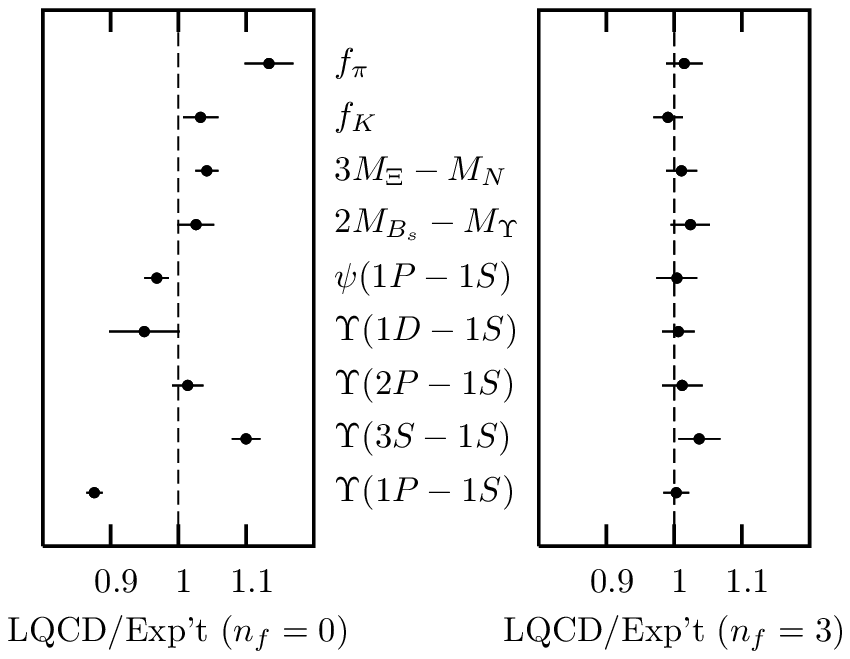}
\end{center}
\caption[]{
Heavy--meson mass splitting (together with some light hadron quantities)
taken from \cite{breakthrough}.
}
\label{fig:heavysplit}
\end{figure}

We reproduce, in Fig. \ref{fig:heavysplit} (taken from \cite{breakthrough})
a graph showing the ratio of lattice prediction to experiment
for some heavy--meson mass splittings. As can be seen, the lattice results
in the full theory (right--hand plot) are within $1\sigma$ of their
experimental values.


\section{Quantifying Quenching effects}
\label{sec:diff}


\subsection{Hadron Spectrum}

In Sects. \ref{sec:quenched} and \ref{sec:full}, we have outlined
some results for the hadronic spectrum for both the quenched
approximation and full QCD.
Comparing these results we note firstly that quenched results are
generally within 10\% of their experimental value for a wide variety
of quantities.
This is an unexpectedly good level of agreement which will be discussed
later in this section.

Studying the light--meson sector, we note that the full theory is able
to accurately reproduce the $K,K^\ast$ and $\phi$ masses to within
around 1\%, far better than the quenched theory (see lower panel of
Fig. \ref{fig:cppacs}).
We note, however, that the ``chiral'' slope,
defined via the $J-$parameter, (\ref{eq:J}), is still several
$\sigma$ away from its experimental value at the simulated values of
$m_{sea}$ (see Fig. \ref{fig:ukqcd_j}), and that the chiral extrapolation,
$m_{sea} \rightarrow m_{u,d}$ is required to make contact with experiment.

This situation is mirrored in the baryonic spectrum.
Figure \ref{fig:cppacs} shows the remarkable prediction from the
CP-PACS collaboration of eight baryonic masses.
In general terms, the agreement between theory and experiment is
enhanced when the quenched approximation is removed.
Note also that there is no discrepancy in the full theory between
predictions using the $K$ and $\phi$ mesons to set the strange quark mass.
The same is not true in the quenched data (see Fig. \ref{fig:cppacs}).
The level of agreement between the full theory prediction and experiment
is most profound for baryons containing the largest strange quark content.
We argued in Sect. \ref{sec:baryons} that this could imply that the
lattice data at the {\em simulated} values of $m_q$ (roughly around
$m_s$) are correct, but that the chiral extrapolation procedure
$m_q \rightarrow m_{u,d}$ is going astray.
As can be seen in Fig. \ref{fig:edplot}, which is roughly the baryonic
equivalent of Fig. \ref{fig:cppacs_meson}, the chiral extrapolation
required to reach the $u,d$ quarks is substantial.

Moving to the heavy--meson sector, we summarised in
Sect. \ref{sec:fullheavy} results from \cite{breakthrough}.
These show excellent agreement between full simulation results and
experiment for a variety of quantities, especially splittings in
the $\upsilon$ spectrum. A corresponding quenched analysis shows
discrepancies of $\sim 10\%$.



\subsection{Why is the Quenched Approximation so good?}

While there are obvious failures in the quenched approximation's
ability to reproduce the real world, it does much better than naive
expectations: one would imagine that removing all $q-\bar{q}$ diagrams
from the theory would have a {\em drastic} effect on the hadron
spectrum.  Figure \ref{fig:cppacs} shows that this is not the case.
Why then does the quenched approximation perform so well?

\begin{figure}[b]
\begin{center}
\includegraphics[width=.8\textwidth]{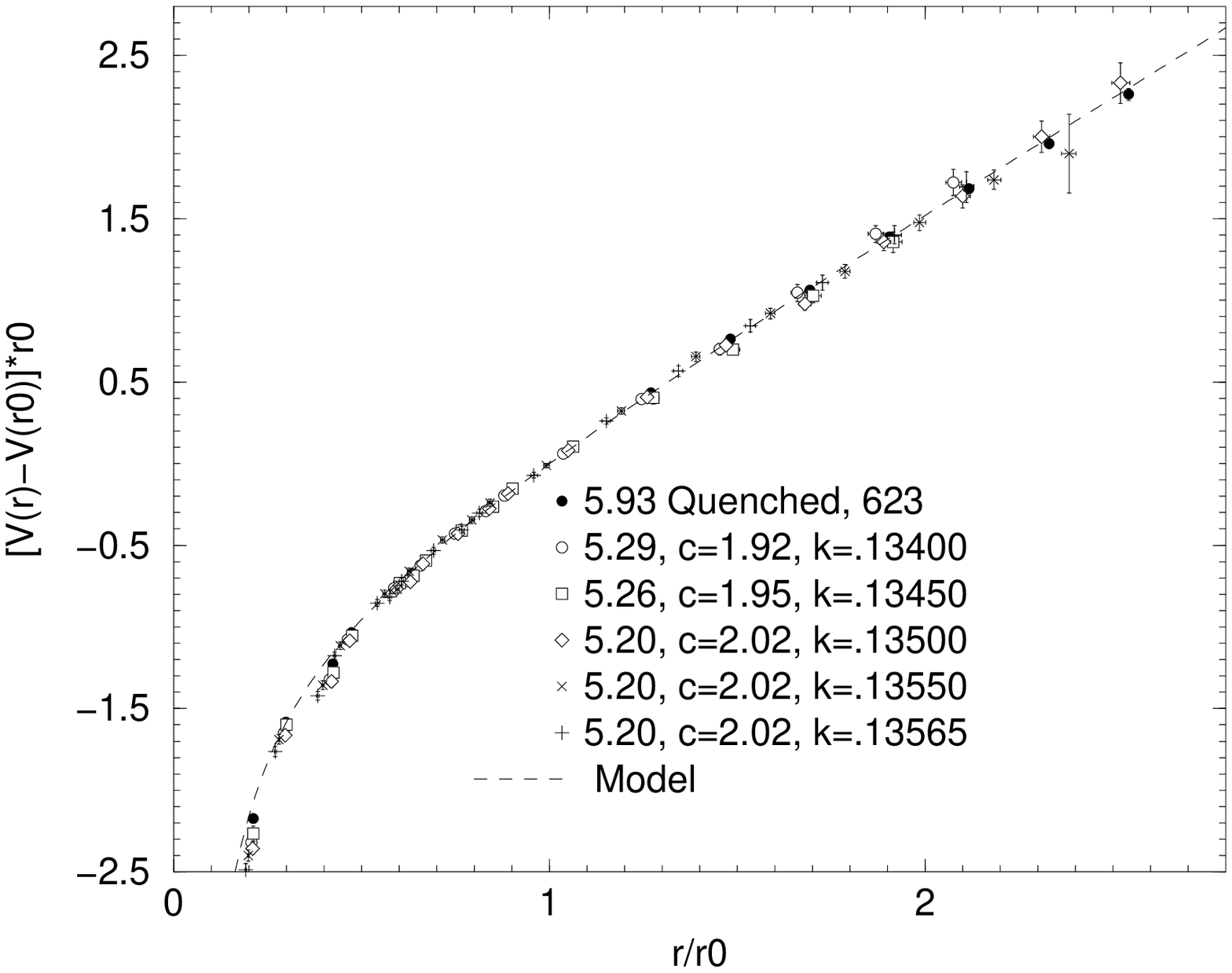}
\end{center}
\caption[]{
The static quark potential from the UKQCD Collaboration \cite{ukqcd}.
The parameters $c$ and $k$ refer to a coefficient of an improvment term
in the action and the sea quark mass parameter respectively.
}
\label{fig:ukqcd_staticpot}
\end{figure}

\begin{figure}[b]
\begin{center}
\includegraphics[width=.8\textwidth]{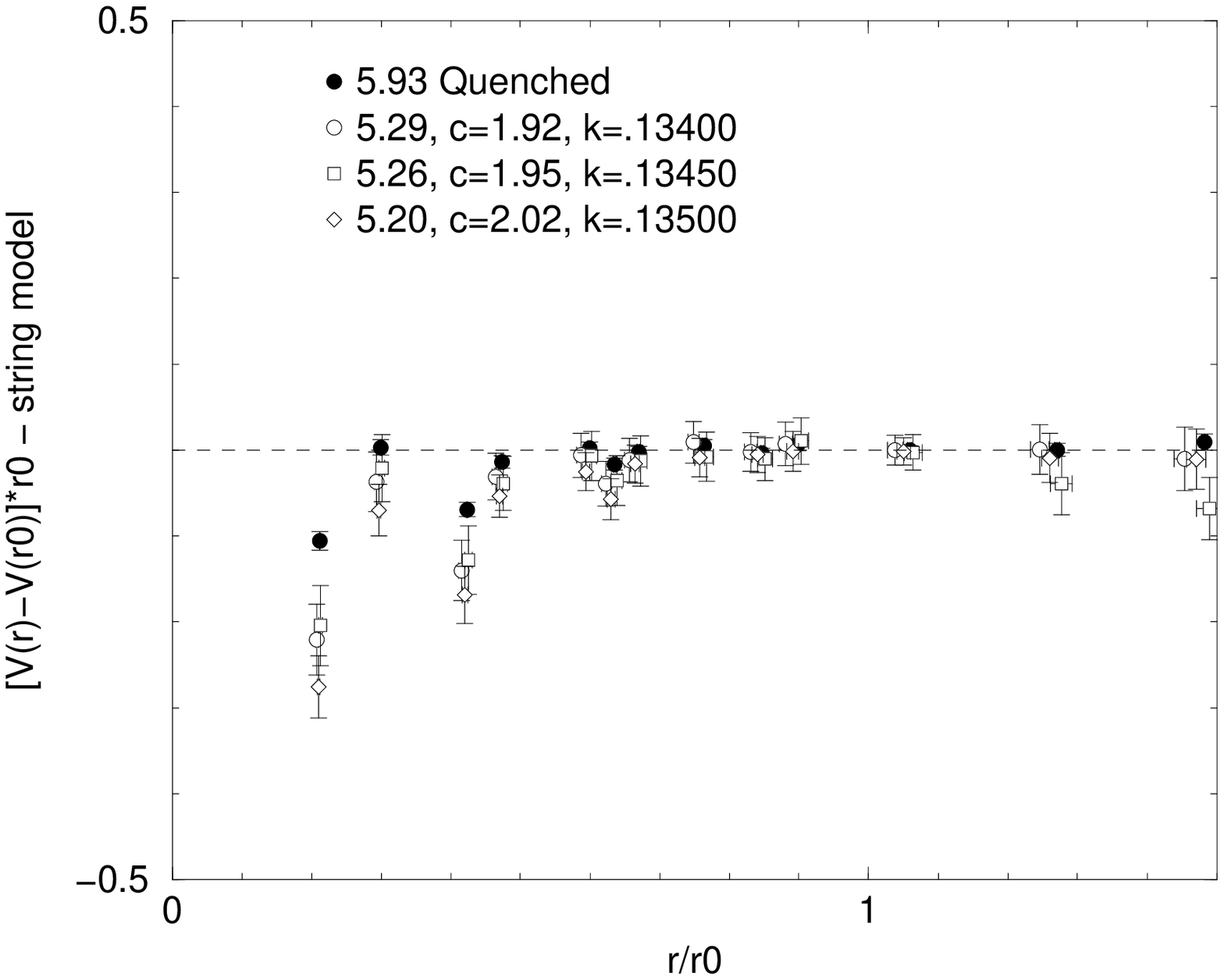}
\end{center}
\caption[]{
The deviation of the static quark potential in
Fig. \ref{fig:ukqcd_staticpot} from the string model \cite{ukqcd}.
}
\label{fig:ukqcd_staticpot_diff}
\end{figure}

One can obtain a handle on this issue by studying the static quark
potential (which is the quantum mechanical potential between two
infinitely heavy quarks).  Figure \ref{fig:ukqcd_staticpot} shows
UKQCD results for this quantity for both the quenched and full theory
\cite{ukqcd}. The curve shown in the graph is the ``string model'',
$V(r) = e/r + \sigma/ r + const$ \cite{luescher}.
Note that the data is {\em defined} to agree in value and slope
exactly at $r=r_0$ (the hadronic scale defined in \cite{sommer}) \cite{ukqcd}.
A close up of the difference between the lattice potential
and the string model at short distances is shown in
Fig. \ref{fig:ukqcd_staticpot_diff}.
As can be seen from
Figs. \ref{fig:ukqcd_staticpot} \& \ref{fig:ukqcd_staticpot_diff},
the discrepancy between the quenched and full theories is negligible
across the whole range of $r$ except at very small distances
where the deviation is discernible, but small.
This implies that only physical quantities
particularly sensitive to this short--distance scale will be affected
by the quenched approximation.  Hadronic states are most sensitive to
\lq\lq medium'' distance scales $r\sim r_0 \approx0.5\,fm$ where (from Fig.
\ref{fig:ukqcd_staticpot}) the two theories' data overlay each other.

Thus the quenched and full theories should agree at the same level as
quarks in QCD can be approximated as moving in a static quark
potential.  This observation presumably has relevance to the age--old
question: Why does the (non-relativistic) quark model perform so well?

It is worth noting that, from Sect. \ref{sec:intro}, the quenched
approximation is defined not just by replacing the quark determinant by
unity, but also by renormalising the coupling $g_0$.
In fact, if you attempt to perform quenched and full simulations
at the {\em same} value of $g_0$, then the lattice spacing, $a$
(or equivalently the cut-off $\sim 1/a$)
will differ by a factor of around four.
This is telling us that the virtual quark loops really are affecting the
dynamics of the simulation.
The apparent contradiction between this fact, and what we have seen
above, i.e. that the quenched approximation reproduces the full theory
(at the $5-10\%$ level) is resolved as follows.
The lattice only actually predicts dimensionless quantities, normally
expressed as, e.g. $M\times a$, where $M$ is some mass.
In this way the lattice is able to predict dimensionless
ratios of physical quantities only, e.g. $M_1/M_2$.
Although switching the quark determinant on and off does directly
affect the lattice spacing, and therefore $Ma$, it seems to have
little effect on the ratio $M_1/M_2$.
In other words the physical prediction from the lattice of $M_1$,
which can be obtained from $M_1/M_2 \times M_2^{expt}$,
doesn't seem to be greatly affected by quenching.
This is telling us something remarkable: for a wide variety of hadronic
masses (and the static quark potential), the removal of virtual quark
loops from the theory can be counter-balanced simply by an adjustment in the
coupling, $g_0$.

There is one final reason why quenching has only a modest affect on
the hadronic spectrum, compared to other physical quantities.
In order to extract a hadron mass from a lattice simulation,
the quantity $\Omega = C(t)$ is calculated (see (\ref{eq:lat}))
where $C(t)$ is a two--point correlation function between hadronic
currents.
In Euclidean space--time, we have
\begin{equation}
C(t) \rightarrow {\cal M}^2 \E^{-Mt} \;\;\;\;
            \mbox{as} \;\;\;\; t\rightarrow \infty
\label{eq:Ct}
\end{equation}
where $\;{\cal M}\;$ is a matrix element between the vacuum and the
hadronic ground state. Lets assume that we are performing a quenched
calculation of $C(t)$ and that it has a relative error of $\varepsilon$ due to
this quenched approximation, i.e.
\begin{equation}
C(t)^Q = C(t)^{full} ( 1 + \varepsilon ),
\label{eq:CtQ}
\end{equation}
where $C(t)^{Q,full}$ are the quenched and full correlation functions
respectively. Because the mass, $M$, appears in the
argument of the exponential, a relatively small adjustment in $M$
can mop up the quenching error $\varepsilon$,
whereas a larger relative change would required of the matrix
element, ${\cal M}$.

Obviously this analysis is a little simplistic but it does
illustrate that we can expect quenching errors in matrix elements
(such as decay constants) to be larger than in masses.


\section{Conclusions}

Lattice QCD is an approach to solving field theories, such as QCD,
which involves no model assumptions.
Given a fast enough computer, the lattice can be used to solve QCD
on any finite volume and with any non-zero quark mass resulting in an absolute
theoretical prediction of QCD.
However, in order to make the problem tractable on current computers,
certain parameters of QCD need to take non-physical values
(see Table \ref{tab:approx}).
The parameter under study in this chapter is the number of quark
flavours in the vacuum, $N_f$.
Setting $N_f = 0$ is called the quenched approximation and
corresponds to ignoring virtual $q-\bar{q}$ pairs in the vacuum.
The approximation $N_f=0$ is seemingly a particularly brutal approximation
and, furthermore, there is little theoretical guidance as to its effect.
Thus we are usually forced to \lq\lq measure''
its effect {\em a posteriori} by analysing data from lattice simulations.

In this chapter we have studied the hadronic spectrum with and without
the quenched approximation, in particular light mesons and baryons,
and heavy--mesons.
By comparing quenched data with experimental masses,
we have shown that quenching effects in the light--hadron spectrum
are relatively small ($5-10\%$), with a slightly larger discrepancy
in the heavy--meson spectrum.
It is only recently that full QCD simulations have
been able to produce data with statistical and systematic errors
beneath this level.
With this new generation of data, we are now
able to state that full QCD lattice results have better agreement
with experiment than quenched results.

We have outlined some reasons why the quenched approximation is
so relatively successful, and we have found evidence that the
chiral extrapolation techniques currently being used in full
QCD simulations require further consideration.

In the future, more precise calculations with, and without the quenched
approximation will surely enhance our understanding of the underlying
physics of QCD.



\section*{Acknowledgements}

The author would like to thank the CSSM in Adelaide for their kind
hospitality.




\end{document}